\UseRawInputEncoding
\documentclass[
 reprint,
superscriptaddress,
 amsmath,amssymb,
 aps,
]{revtex4-1}

\usepackage{graphicx}
\usepackage{bm}
\usepackage{hyperref}
\usepackage{braket}
\usepackage{commath}
\usepackage{color}
\usepackage{subfigure}
\usepackage{epstopdf}
\usepackage{epsfig}
\usepackage{amsmath}
\usepackage{mathtools}
\usepackage{dcolumn}
\usepackage{url}
\usepackage{hyperref}
\hypersetup{%
        plainpages=true,
        breaklinks=true,
        hypertexnames=false,
        pageanchor=true,
        colorlinks=true,
        linkcolor={blue},
        citecolor={magenta},
        urlcolor={blue},
        anchorcolor={black}
      }

\usepackage{mathrsfs,bm,amsthm,amsfonts,xspace,float,bookmark}
\usepackage{textcomp}
\usepackage{tabu}
\usepackage{bbold}
\usepackage{xspace}
\usepackage{float}
\usepackage[inline]{enumitem}
\AtBeginDocument{%
    \newwrite\bibnotes
    \def\bibnotesext{Notes.bib}
    \immediate\openout\bibnotes=\jobname\bibnotesext
    \immediate\write\bibnotes{@CONTROL{REVTEX41Control}}
    \immediate\write\bibnotes{@CONTROL{%
    apsrev41Control,author="08",editor="1",pages="1",title="0",year="1"}}
     \if@filesw
     \immediate\write\@auxout{\string\citation{apsrev41Control}}%
    \fi
}%

\makeatletter
\newcommand*{\closeindex}[1]{_{\mkern-4.5mu#1}}
\DeclareRobustCommand{\power}{\textrm{\textit{\textsf{P}}}\@ifnextchar_{\expandafter\closeindex\@gobble}{}}
\DeclareRobustCommand{\gain}{\textrm{\textit{\textsf{G}}}\@ifnextchar_{\expandafter\closeindex\@gobble}{}}
\makeatother
\renewcommand{\thesection}{\Roman{section}}
\usepackage{units}
\newcommand{\Gama}[1]{{{\Gamma_{\rm#1}}}}
\newcommand{\Gamaa}[2]{{\it{\Gamma_{\rm#1}^{\rm#2}}}}
\newcommand{\Omga}[1]{{{\Omega_{\rm#1}}}}
\newcommand{\Delt}{{\it{\Delta}}}

\usepackage{amsmath}

\newcommand{\be}{\begin{equation}}
\newcommand{\ee}{\end{equation}}
\newcommand{\bea}{\begin{eqnarray}}
\newcommand{\eea}{\end{eqnarray}}

\newcommand{\beq}{\begin{eqnarray}}
\newcommand{\eeq}{\end{eqnarray}}
\renewcommand{\eqref}[1]{\mbox{Eq.~(\ref{#1})}}

\newcommand{\ex}[1]{\langle #1 \rangle}

\begin{document}

\title{Nonequilibrium heat transport and work with a single artificial atom coupled to a waveguide: emission without external driving}

\author{Yong Lu}
\email[e-mail:]{kdluyong@outlook.com}
\affiliation{Microtechnology and Nanoscience, MC2, Chalmers University of Technology, SE-412 96
G\"oteborg, Sweden}
\author{Neill Lambert}
\email[e-mail:]{nwlambert@gmail.com}
\affiliation{Theoretical Quantum Physics Laboratory, RIKEN Cluster for Pioneering Research, Wako-shi, Saitama 351-0198, Japan}
\author{Anton Frisk Kockum}
\affiliation{Microtechnology and Nanoscience, MC2, Chalmers University of Technology, SE-412 96
G\"oteborg, Sweden}
\author{Ken Funo}
\affiliation{Theoretical Quantum Physics Laboratory, RIKEN Cluster for Pioneering Research, Wako-shi, Saitama 351-0198, Japan}
\author{Andreas Bengtsson}
\affiliation{Microtechnology and Nanoscience, MC2, Chalmers University of Technology, SE-412 96
G\"oteborg, Sweden}
\author{Simone Gasparinetti}
\affiliation{Microtechnology and Nanoscience, MC2, Chalmers University of Technology, SE-412 96
G\"oteborg, Sweden}

\author{Franco Nori}
\affiliation{Theoretical Quantum Physics Laboratory, RIKEN Cluster for Pioneering Research, Wako-shi, Saitama 351-0198, Japan}
\affiliation{Department of Physics, The University of Michigan, Ann Arbor, 48109-1040 Michigan, USA}
\author{Per Delsing}
\affiliation{Microtechnology and Nanoscience, MC2, Chalmers University of Technology, SE-412 96
G\"oteborg, Sweden}

\date{\today}%

\begin{abstract}
 We observe the continuous emission of photons into a waveguide from a superconducting qubit without the application of an external drive. To explain this observation, we build a two-bath model where the qubit couples simultaneously to a cold bath (the waveguide) and a hot bath (a secondary environment). Our results show that the thermal-photon occupation of the hot bath is up to $0.14$ photons, 35 times larger than the cold waveguide, leading to nonequilibrium heat transport with  a power of up to 132 zW, as estimated from the qubit emission spectrum. By adding more isolation between the sample output and the first cold amplifier in the output line, the heat transport is strongly suppressed. Our interpretation is that the hot bath may arise from active two-level systems being excited by noise from the output line. We also apply a coherent drive, and use the waveguide to measure thermodynamic work and heat, suggesting waveguide spectroscopy is a useful means to study quantum heat engines and refrigerators. Finally, based on the theoretical model, we propose how a similar setup can be used as a noise spectrometer which provides a new solution for calibrating the background noise of hybrid quantum systems. 
 \end{abstract}

\keywords{Suggested keywords}
\maketitle
\section{Introduction}


Over the past 20 years, superconducting qubit coherence times have increased from less than \unit[1]{ns} to more than \unit[1]{ms}~\cite{gu2017microwave, kjaergaard2020superconducting, place2020new, somoroff2021millisecond}. Many studies have shown that such coherence times are limited by two-level systems (TLSs)~\cite{lisenfeld2015observation, Weides2019correlating, muller2019towards} and that quasiparticles can also contribute~\cite{gustavsson2016suppressing,Catelani2011quasiparticle,wang2014measurement,de2020two}.  
Moreover, excessive thermal population of a qubit, arising from nonequilibrium quasiparticles, has been observed extensively, with effective temperatures in the range \unit[30--200]{mK}~\cite{jin2015, Wenner2013excitation, DiCarlo2012Feedback}. More recently, ionizing radiation due to high-energy cosmic rays and radioactive decay has been shown to reduce qubit coherence~\cite{cardani2020reducing, vepsalainen2020impact, pop2018loss}. Therefore, isolating superconducting qubits from all possible sources of noise will be crucial for realizing fault-tolerant superconducting quantum computers~\cite{vepsalainen2020impact,mcewen2021removing, Egger2018}.

In addition, in the field of quantum thermodynamics, alongside the study of quantum heat engines~\cite{rossnagel2016single, peterson2019experimental, Ono2020Heat,quan2007quantum} and refrigerators~\cite{maslennikov2019quantum, xu2020radiative}, the dynamical control of heat flow, mediated by phonons or photons, has been investigated in solid-state circuits~\cite{schoonveld2000coulomb}, nanostructures~\cite{schwab2000measurement, chang2006solid}, and superconducting circuits~\cite{partanen2016quantum,meschke2006single}. 
{More recently, the potential for significant quantum effects was introduced into such studies in the form of a superconducting qubit coupled to cavities~\cite{quan2006maxwell,naghiloo2020heat, Cottet7561, senior2020heat, cherubim2019non}.} In particular, a heat valve for energy transfer between two artificial heat baths constructed from resonators has been realized, where the transferred power was derived from the temperature difference between the two baths which is measured by thermometers~\cite{ronzani2018tunable, senior2020heat}. The direct coupling of a qubit to a waveguide, a setting known as quantum electrodynamics (QED)~\cite{w2017roy,kockum2019quantum} has been suggested as a platform for studies in thermodynamics~\cite{Monsel2020Work}, but has not yet been experimentally explored.

In this work, we first observe the emission from a superconducting qubit into a cold waveguide due to thermal excitation of the qubit by a hot bath, and investigate the potential source of this additional thermal noise. We then show how the qubit emission spectrum and reflectivity can be used to measure the heat and work rates, in analogy to the heat valves demonstrated with engineered environments~\cite{ronzani2018tunable, senior2020heat}.

\begin{figure}[tbph]
\includegraphics[width=1\linewidth]{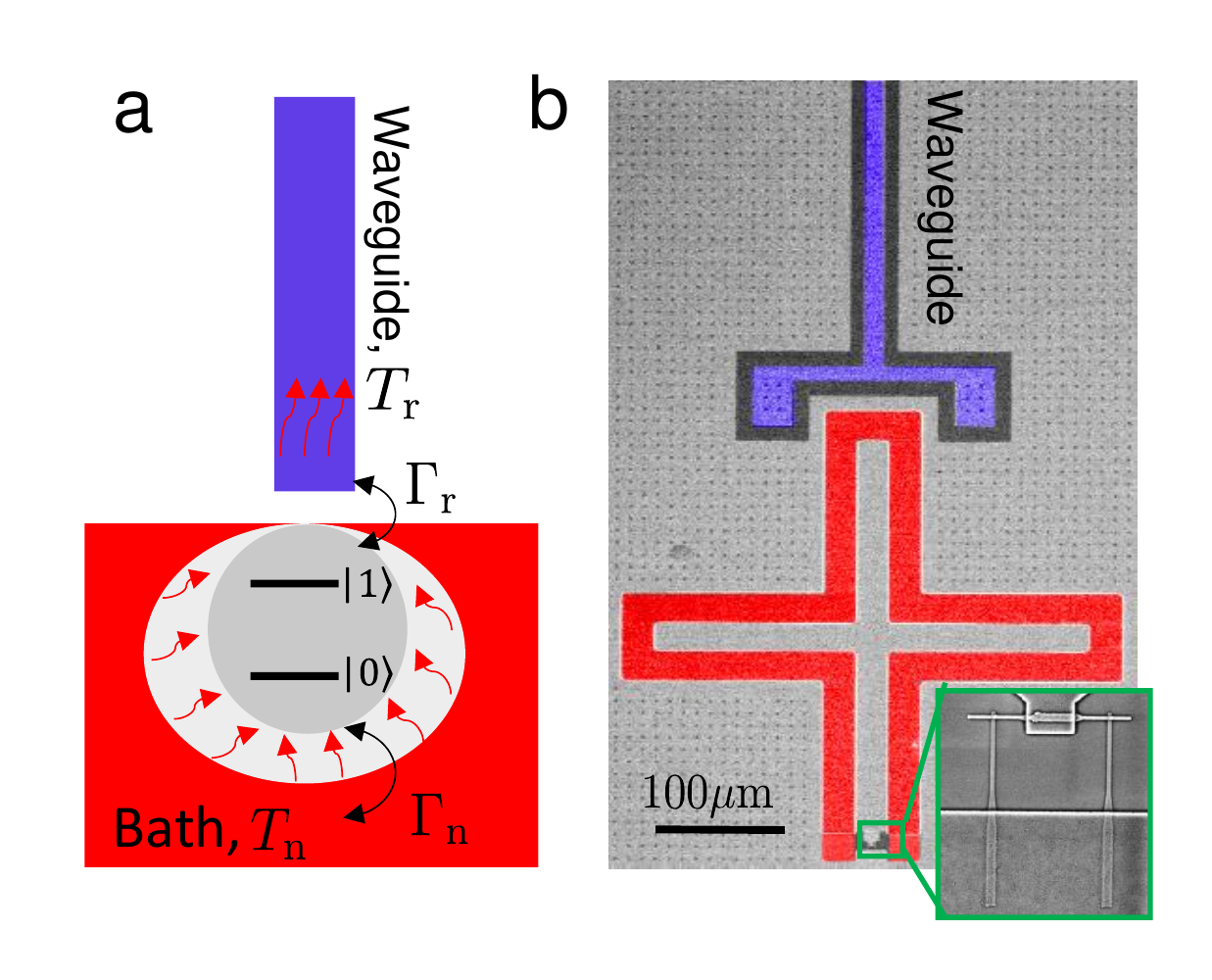}
\caption{\textsf{{\bf{\textsf{Sketch of the experimental setup}}.}
{\bf{a}}, A schematic showing how the two-level artificial atom ($|0\rangle$ and $|1\rangle$ are the corresponding ground and excited states of the atom) absorbs energy from the hot bath surrounding the atom. Therefore, the excited atom emits photons to the cold waveguide. We have $\Gama{r}$ and $\Gama{n}$ as the emission rates of the atom into the waveguide and the non-radiative bath, respectively.
{\bf{b}}, False-color micrograph of the superconducting circuit realizing the setup in panel {\bf{a}}: A transmon qubit consisting of a cross-shaped superconducting island shunted by a superconducting quantum interference device (SQUID) capacitively coupled to a microwave coplanar waveguide (blue). The inset shows a close-up of the SQUID. [See more details on the measurement setup in the Methods section]
}
}
\label{setupcartoon}
\end{figure}
\section{Results}
\begin{figure*}[tbph]
\includegraphics[width=0.9\linewidth]{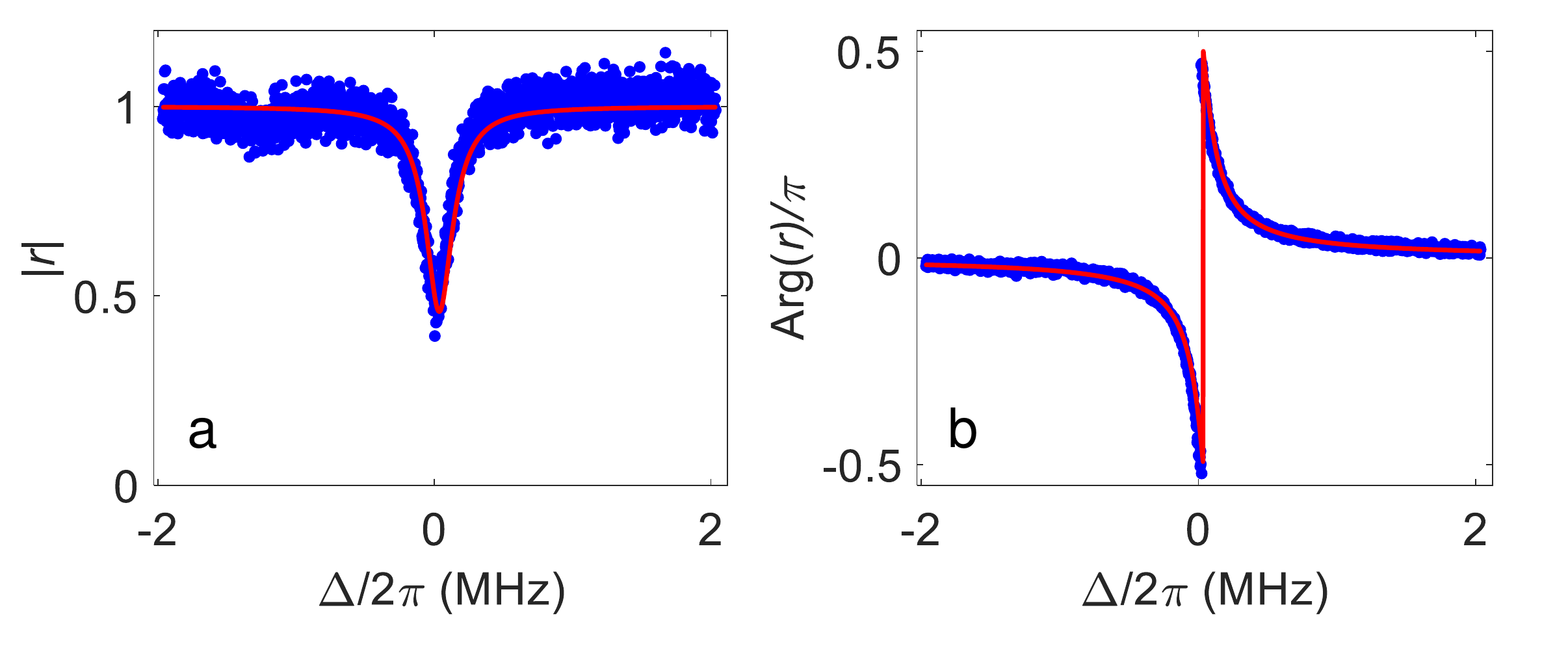}
\caption{\textsf{{\bf{\textsf{Weak-power spectroscopy under a coherent drive}}.}
{\bf{a}}, Magnitude and {\bf{b}}, phase response of the reflection coefficient $r$ as a function of the probe detuning. Blue dots are the experimental data with the red solid curves calculated from Eq.~(\ref{reflection}).
}
}
\label{spectroscopy}
\end{figure*}

{\bf{Theoretical model}}. To model our experiment, we introduce a two-bath model where the qubit is simultaneously coupled to a cold (radiative) waveguide bath and a hot auxiliary (nonradiative) environment. We also include a coherent drive, at frequency $\omega_{\rm{p}}$, through the waveguide in the system. Thus, we have a qubit and two bosonic environments described by the Hamiltonian 
\beq\label{H1}
H = H_{\rm{q}} + H_{\rm{r}} + H_{\rm{n}}
\eeq
with
\beq\label{H2}
\frac{H_{\rm{q}}}{\hbar} = -\frac{\Delta}{2}\sigma_z + \frac{\Omega}{2}\sigma_x,
\eeq
where the qubit Hamiltonian is in the rotating frame with $\Delta = \omega_{\rm{p}}-\omega_{01}$ and $\Omega$ as the probe strength. The radiative (waveguide) Hamiltonian $H_{\rm{r}}$ and the nonradiative bath Hamiltonian $H_{\rm{n}}$, including couplings to the qubit (under the rotating-wave-approximation), are given by
\beq\label{H3}
\frac{H_{\emph{i}}}{\hbar}=\sum_{\emph{k}} \omega_{k,i} a_{k,i}^{\dagger}a_{k,i} + \sum_k g_{k,i} \left(\sigma_- a_{{k,i}}^{\dagger} + \sigma_+a_{{k,i}} \right),
\eeq
where $g_{{k,i}}$ is the coupling strength to mode ${k}$ in environment $i\in\{\rm{r}\,,\rm{n}\}$ at frequency $\omega_{{k,i}}$, and $a_{k,i} (a_{k,i}^{\dagger})$ is the corresponding annihilation (creation) operator of the mode.
Hereafter, we treat both environments under the standard Born-Markov secular (BMS) approximations~\cite{breuer2002theory}, and use this model to fit the single-tone spectroscopy and power spectrum. Under the BMS approximation [Supplementary material S2], we define $\Gama{r}$ and $\Gama{n}$ as the radiative and the nonradiative decay rates, respectively and $n_{\rm{r}}$ and $n_{\rm{n}}$ as the thermal occupations  at temperatures $T_{\rm{r}}$ and $T_{\rm{n}}$, respectively, for the two baths. A corresponding explanatory diagram is shown in Fig.~\ref{setupcartoon}(a).

Our device is a frequency-tunable transmon-type artificial atom~\cite{Koch2007charge} coupled to a 1D semi-infinite waveguide, terminated by an open end, acting as a mirror in Fig.~\ref{setupcartoon}(b)~\cite{hoi2015probing,amplification2018Hoi, lu2021propagating, scigliuzzo2020primary,lin2020deterministic}. We operate the qubit  at the \unit[10]{mK} stage of the mixing chamber of the dilution refrigerator at its maximum frequency $\omega_{01}/2\pi \approx \unit[5.5]{GHz}$, where the qubit is flux-insensitive to first order, thus minimizing the pure dephasing rate.

{\bf{Qubit spectroscopy}}. We first characterize our qubit by single-tone spectroscopy where we send a weak coherent tone to the qubit ($\Omga{}\ll\Gama{r}$) and then measure the coherent reflection coefficient $r$. Sweeping the probe frequency across the resonance of the qubit we observe that the reflection coefficient shows a dip in the amplitude [Fig.~\ref{spectroscopy}(a)] and a $\pi$ phase shift [Fig.~\ref{spectroscopy}(b)].  According to input-output theory with two baths under a weak probe, the reflection coefficient is
\beq
r=1 - i\Gama{r}\frac{(1-2\Gama{+}/\Gama{1} )}{\Delt+i\Gama{2}},
\label{reflection}
\eeq
where $\Gama{+}=n_{\rm{n}}\Gama{n}+n_{\rm{r}}\Gama{r}$, $\Gama{1}=(1+2n_{\rm{n}})\Gama{n}+(1+2n_{\rm{r}})\Gama{r}$, and $\Gama{2}=\Gama{1}/2+\Gama{\phi}$. The pure dephasing rate $\Gama{\phi}$ can also be affected by the thermal excitation of the third level of the transmon~[Supplementary Material S6].
{By using Eq.~(\ref{reflection}), we calculate the theoretical solid curves with $\Gama{r}(1-2\Gama{+}/\Gama{1})/2\pi=\unit[214]{kHz}$ and $\Gama{2}/2\pi=\unit[147]{kHz}$ shown in Fig.~\ref{spectroscopy}.}
\begin{figure*}[tbph]
\includegraphics[width=0.9\linewidth]{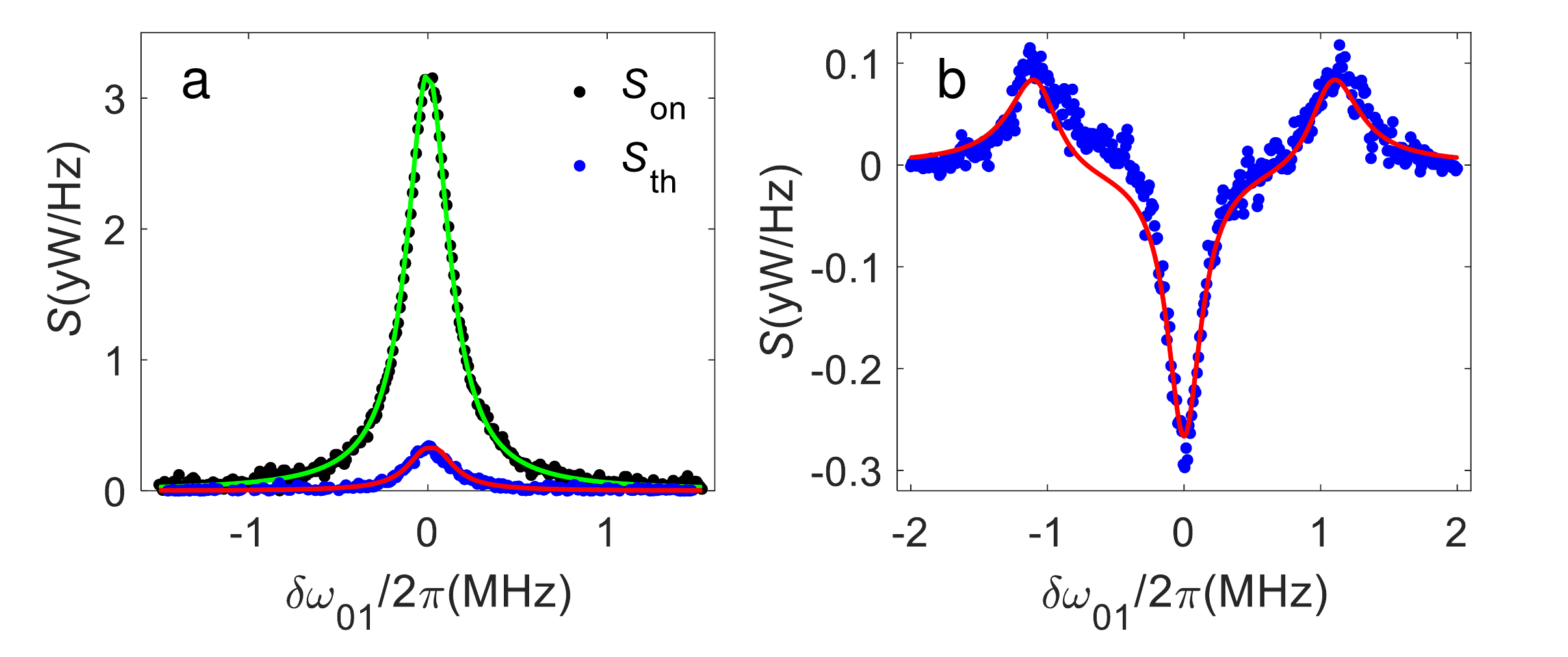}
\caption{\textsf{{\bf{\textsf{Thermal spectrum}}.}
{\bf{a}}, Power spectral density (PSD) of the output field where the results with and without a strong drive are $S_{\rm{on}}$ and $S_{\rm{th}}$, respectively. Units are yW/Hz=$10^{-24}$ W/Hz and $S=2\pi S(\omega)$. The solid curves are the corresponding fits to Eq.~(\ref{swon}) and Eq.~(\ref{thermalspectrum}). The Rabi frequency used for $S_{\rm{on}}$ on the $|0\rangle\leftrightarrow|1\rangle$ transition is $\Omga{}/2\pi\approx\unit[8.8]{MHz}$.
{\bf{b}}, Power spectral density from Autler-Townes splitting. The solid curve is calculated from theory. In both {\bf{a}} and {\bf{b}}, dots are from the experimental data. The Rabi frequency used for the drive on the $|1\rangle\leftrightarrow|2\rangle$ transition is $\Omga{}/2\pi\approx\unit[1.5]{MHz}$. {Note that the PSD in {\bf{a}} is subtracted by the background when tuning the qubit frequency away through the external magnetic flux, whereas in {\bf{b}} the PSD is the difference between the thermal spectra with and without the drive on the $|1\rangle\leftrightarrow|2\rangle$ transition.}
}
}
\label{spectrum}
\end{figure*}

{\bf{Qubit power spectrum density}}. Without sending any signal from room temperature to the sample, we observe a power spectral density (PSD) emitted from the qubit into the waveguide, shown as blue dots in Fig.~\ref{spectrum}(a) [more measurement details are presented in the Methods section]. The Lorentzian shape centered at the qubit frequency indicates that the environment surrounding the qubit is hotter than the waveguide, and is causing heat flow into the waveguide through the qubit. In order to justify this statement, and fit the data, we use standard input-output theory for the output of the transmission line $b_{\mathrm{out}}(t) =   b_{\mathrm{in}}(t) - i\sqrt{\Gama{r}} \sigma_-(t) $, where $b_{\mathrm{in}}(t) = f_{\mathrm{in}}(t) + \frac{\Omega}{2\sqrt{\Gama{1}}}$ includes thermal noise $f_{\mathrm{in}}(t)$ and the coherent drive $\frac{\Omega}{2\sqrt{\Gama{1}}}$, for the cases when such applied. It is necessary to take into account the fact that the thermal input $f_{\mathrm{in}}(t)$ to the waveguide  can become correlated with the qubit emission operator $ \sigma_-(t)$ to evaluate the correct output spectrum~\cite{Gardiner04},  given by [Supplementary Material S5]
\beq
S(\omega) =\frac{\hbar\omega_{01}}{2\pi} \int_{-\infty}^{\infty} dt e^{-i\omega t} \langle  b_{\mathrm{out}}^{\dagger}(t) b_{\mathrm{out}}(t')\rangle_{(t' \rightarrow \infty)}
\eeq
with
\beq\label{outp}
\langle  b_{\mathrm{out}}^{\dagger}(t) b_{\mathrm{out}}(t')\rangle &=& \Gama{r}\left(n_r+1\right)\langle \sigma_+(t) \sigma_-(t')\rangle\\
&-& \Gama{r} n_r
\langle  \sigma_-(t') \sigma_+(t)\rangle + \langle f_{\mathrm{in}}^{\dagger}(t)f_{\mathrm{in}}(t')\rangle \nonumber\\ &-& \frac{i\Omega^*}{2}\langle \sigma_-(t')\rangle + \frac{i\Omega}{2} \langle\sigma_+(t)\rangle +  \frac{|\Omega|^2}{4\Gama{r}}. \nonumber
\eeq

\begin{figure*}[tbph]
\includegraphics[width=1\linewidth]{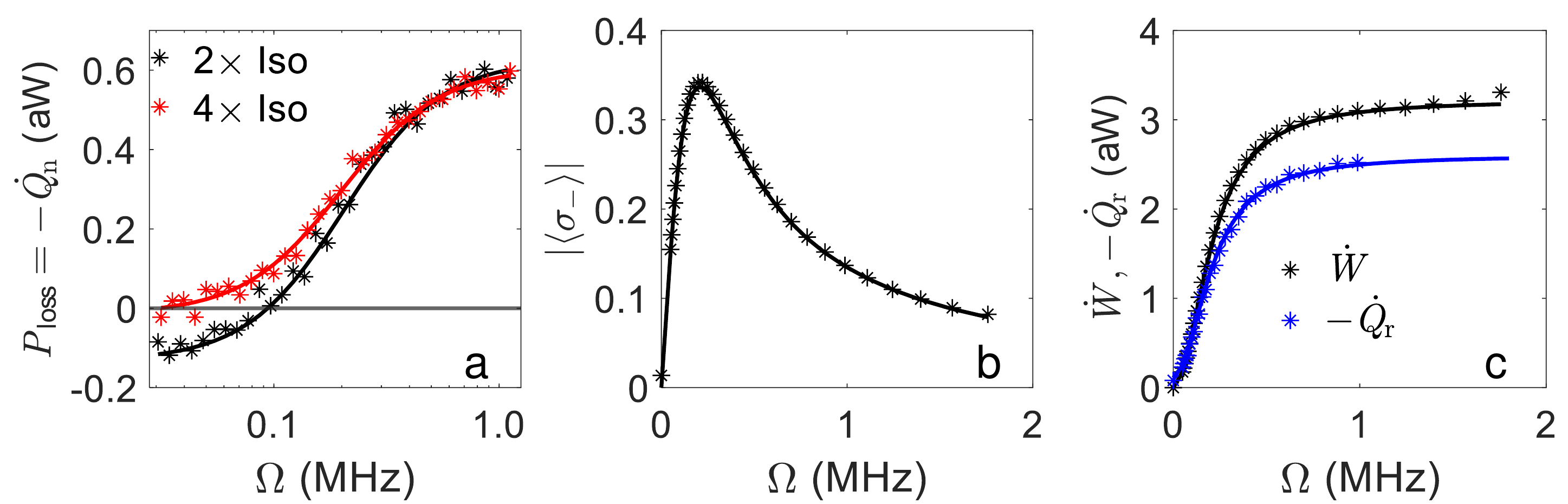}
\caption{\textsf{{\bf{\textsf{Power loss, work, and heat}}.}
{\bf{a}}, The lost power $P_{\rm{loss}}$ as a function of the Rabi frequency ($\Omega$) of the drive with two and four isolators.
{\bf{b}}, The magnitude of the qubit emission as a function of $\Omega$ of the drive with two isolators.
{\bf{c}}, Work ($\dot{W}$) performed and heat ($\dot{Q}$) generated by the drive as a function of $\Omega$. In all panels, {the stars are the experimental data and the solid curves are from theory, where the blue stars in (c) are from the difference between the values of the black stars in {\bf{c}} and {\bf{a}} according to $\dot{Q_{\rm{r}}}=P_{\rm{loss}}-\dot{W}$.}
}
}
\label{powerloss}
\end{figure*}

This expression is easily generalized to  include more levels within the artificial atom. Importantly, if we define the Fourier transforms of the first two terms of Eq.~(\ref{outp}), $s_q^+(\omega) = \int_{-\infty}^{\infty}dt \langle \sigma_+(t) \sigma_-(t')\rangle e^{-i\omega t}$, $s_q^-(\omega) = \int_{-\infty}^{\infty}dt \langle \sigma_-(t
') \sigma_+(t)\rangle e^{-i\omega t}$, then the limited detailed balance that arises for a qubit coupled to a single BMS environment, e.g., the waveguide, tells us that $s_q^-(\omega) =e^{\beta_r \omega_{01}}s_q^+(\omega)$, where $\beta_r = 1/k_{\mathrm{B}} T_r$, and $k_{\mathrm{B}}$ is the Boltzmann constant. Hence, in equilibrium situations (no drive and no additional environment $H_{\mathrm{n}}$), the first two terms in Eq.~(\ref{outp}) cancel, resulting in $S(\omega)=0$.  

Explicitly evaluating the correlation functions for the qubit using our two-bath master equation [omitting both delta-function coherent contributions~\cite{koshino} and the background thermal black-body spectrum of the transmission line in \eqref{outp}], we find the thermal spectral density output ($\Omega =0$) is given by
\beq
S_{\rm{th}}(\omega) =\hbar\omega_{01}\frac{\Gama{r}}{2\pi}\frac{2\Gama{2}\Gama{n}\Delta n/\Gama{1}}{\delta\omega_{01}^2 + \Gama{2}^2},
\label{thermalspectrum}
\eeq
where $\Delta n=n_{\rm{n}}-n_{\rm{r}}$, and $\delta\omega_{01}=\omega-\omega_{01}$. In Eq.~\ref{thermalspectrum}, we find that $S_{\rm{th}}(\omega) =0$ if $n_{\rm{n}}=n_{\rm{r}}$, as expected from detailed balance.  Without the additional nonradiative bath to thermally excite the qubit, we would not observe a thermal response (even in the presence of thermal input to the transmission line). As shown in Fig.~\ref{spectrum}(a), $S_{\rm{th}}(\omega)\neq0$ in the experiment, which implies that the waveguide and the bosonic bath surrounding the qubit are not in equilibrium. Fitting Eq.~(\ref{thermalspectrum}) to the data in Fig.~\ref{spectrum}(a) gives $\Gama{r}\Gama{n}\Delta n/(2\pi\Gama{1})=\unit[5.6\pm0.2]{kHz}$ corresponding to a transported power $\int_{-\infty}^{\infty}S_{\rm{th}}(\omega) d\omega=\unit[132\pm5]{zW}$
with the given $\Gama{2}$ value from the reflection-coefficient measurement.

We also send a strong drive on resonance with the qubit ($\Omga{}/2\pi\approx\unit[8.8]{MHz}\gg\Gama{1}$), where such a large Rabi frequency ensures that there is no overlap between the sidebands and the center of the Mollow-triplet PSD~\cite{astafiev2010resonance,lu2021characterizing}. For the middle peak from the elastic scattering, we have
\beq
S_{\rm{on}}(\omega) = \hbar\omega_{01}\frac{\Gama{r}}{2\pi}\frac{\Gama{2}/2 }{\delta\omega_{01}^2 + \Gama{2}^2}.\label{swon}
\eeq

By fitting the data with the drive on [black dots in Fig.~\ref{spectrum}(a)], we obtain $\Gama{r}/2\pi=\unit[227\pm 4]{kHz}$ and  $\Gama{2}/2\pi=\unit[143\pm 4]{kHz}$. Compared to Eq.~(\ref{thermalspectrum}), the height of this spectrum is mostly independent of the thermal occupation difference $\Delta n$, due to the saturation of the first excited state.

{\bf{Heat transport and work}}. Another straightforward way to evaluate the heat transport between the two baths is to measure the lost power $P_{{\rm{loss}}}$ due to the nonradiative decay. Shown as black in Fig.~\ref{powerloss}(a), we find that when the coherent drive is weak, the thermal noise dominates the qubit dynamics. Therefore, the qubit absorbs photons from the hot bath and then decays into the cold waveguide with a higher probability compared to the overall emission back into the hot bath, resulting in a negative lost power. When we increase the photon-flux occupation in the waveguide by increasing the drive intensity, the absolute value of $P_{{\rm{loss}}}$ is reduced. At $\Omga{Rabi}/2\pi\approx\unit[95]{kHz}$, we see that equilibrium is reached, with zero lost power. However, beyond this value, the waveguide occupation becomes larger than the non-radiative environment. Thus, the qubit is excited by the photons in the waveguide and then dissipates into both environments via nonradiative decay. By increasing the drive intensity further, the qubit is saturated, with the population reaching the maximum value of 0.5 in the steady state. Therefore, $P_{{\rm{loss}}}\approx\hbar\omega_{01}\frac{\Gama{n}}{2}$. Theoretically, the lost power is given by
$P_{{\rm{loss}}}=\hbar\omega_{01}(\langle  b_{\mathrm{in}}^{\dagger} b_{\mathrm{in}}\rangle- \langle  b_{\mathrm{out}}^{\dagger} b_{\mathrm{out}}\rangle) $ as
\beq
P_{{\rm{loss}}}=\hbar\omega_{01}\frac{\Gama{n}}{2}\frac{\Omga{}^2-2\Gama{2}\Gama{r}\Delta n}{\Omga{}^2+\Gama{2}\Gama{1}},
\label{powerlosseq}
\eeq
By fitting Eq.~(\ref{powerlosseq}) to the data using the values of $\Gama{2}$ and $\Gama{r}$ extracted from the fit of Eq.~(\ref{swon}), with $\Omega$ calibrated by the Mollow-triplet [black solid curve], {we obtain $\Gama{n}/2\pi = \unit[55\pm3]{kHz}$ with $\Delta n=0.135$ photons and $\Gama{1}/2\pi=\unit[299]{kHz}$}. The transported power from the hot bath is about \unit[132]{zW} which is consistent with the integral of the thermal spectrum in Fig.~\ref{spectrum}(a).

To obtain the temperature of the waveguide and the bath separately, we combine the value of $\Delta n$ with the result from the reflection-coefficient measurement where we can obtain the thermal population of the qubit {$\rho_{11}^{\rm{th}}=\Gama{+}/\Gama{1}\approx \unit[2.86]{\%}$ from the extracted parameters. Consequently, we obtain $n_{\rm{r}}\approx0.004$, $n_{\rm{q}}\approx0.03$ and $n_{\rm{n}}\approx0.139$ corresponding to $T_{\rm{r}}\approx \unit[50]{mK}$ for the cold waveguide, $T_{\rm{q}}\approx \unit[78]{mK}$ for the qubit and $T_{\rm{n}}\approx \unit[131]{mK}$ for the hot bath, respectively. Therefore, we can obtain the value of $\Gama{1}=(1+2n_{\rm{n}})\Gama{n}+(1+2n_{\rm{r}})\Gama{r}\approx \unit[2\pi\times299\pm7]{kHz}$ which is very close to 2$\Gama{2}$, indicating that the pure dephasing rate $\Gama{\phi}$ is negligible at the flux sweet spot.}

In order to make sure that the environment is not affected when changing the external magnetic flux to tune the qubit away from its maximum frequency, we send a strong drive on the transition between the second and third levels of the transmon, inducing an Autler-Townes splitting~\cite{Autler1955Stark}. Then, we obtain the PSD subtracted from the reference with the drive off, {where we observe two peaks and one dip [blue in Fig.~\ref{spectrum}(b)]. The two peaks are the thermal spectra from the thermal emission of the dressed states when a strong drive is applied on the $|1\rangle\leftrightarrow|2\rangle$ transition. The distance between these two peaks are twice of the corresponding Rabi frequency : $2\Omega\approx2\pi\times\unit[3.0]{MHz}$. When the drive is off, as we discussed, we have a thermal spectrum $S_{\rm{th}}$ at the undressed qubit frequency. Importantly, we have a dip at that frequency when we subtract the data with the drive off from the data with the drive on.} We find that the numerical result from the master equation [red in Fig.~\ref{spectrum}(b)] matches well the data [blue in Fig.~\ref{spectrum}(b)] with parameters $\Gama{r}/2\pi= \unit[227]{kHz}$, $\Gama{n}/2\pi= \unit[55]{kHz}$ and $\Delta n=0.135$. These parameters agree fully with those extracted from the power-loss measurements.

When we drive the qubit, the external field is doing work on this single-atom quantum system. Defining $H_0 = \frac{\hbar\omega_{01}}{2}\sigma_z$ as the bare Hamiltonian which measures the internal energy of the qubit, and $H_1 = \frac{\hbar\Omega}{2}\sigma_x$ as the drive Hamiltonian, the work performed on the qubit by the coherent drive is defined as~\cite{Cottet7561, quantum2013pekola}
\beq
{\dot{W}} = \mathrm{Tr}\left\{-i[H_0, H_1]\rho\right\} = \frac{\hbar\Omega}{2}\ex{\sigma_y}={\hbar\Omega}\Re{[i\ex{\sigma_-}]}
\eeq
Therefore, the work $\dot{W}$ only depends on the qubit coherent emission $\ex{\sigma_-}$ with $\ex{\sigma_-}=i(\ex{b_{\mathrm{out}}}-\ex{b_{\mathrm{in}}})/\sqrt{\Gama{r}}$. In particular, when the drive is on resonance with the qubit [see Supplementary Material Eq.~(S7)], it leads to $\dot{W}= i\hbar{\Omega}\ex{\sigma_-}=\hbar{\Omega}|\ex{\sigma_-}|$. We obtain $\ex{\sigma_-}$ from the values of $\ex{b_{\mathrm{out}}}$ and $\ex{b_{\mathrm{in}}}$ which are measured with and without the qubit by tuning the qubit frequency. The results match well with theory [black in Fig.~\ref{powerloss}(b) and (c)], where the work increases with the drive intensity and then saturates around \unit[3.2]{aW} due to the saturation of the qubit.
\begin{table}
 \caption{{Summary of the system parameters including different qubit decay rates and temperatures of the qubit, waveguide and the hot bath.}} \label{tab:1}
  \centering
\begin{tabular*}{\columnwidth}{  @{\extracolsep{\fill}} c c c c c c c c  @{} }
  \hline
  \hline
  $\Gamma_{\rm{r}}/2\pi$ & $\Gamma_{\rm{n}}/2\pi$  &$\Gamma_1/2\pi$&$\Gamma_2/2\pi$&$T_{\rm{q}}$&$T_{\rm{r}}$&$T_{\rm{n}}$\\
         kHz&kHz&kHz&kHz&mK&mK&mK\\
  \hline
   $227\pm4$ & $55\pm3$ & $299\pm7$ & $143\pm4$ &78 & 50 & 131\\
  \hline
  \hline
\end{tabular*}

\end{table}

Besides the work, the heat from the waveguide and the hot bath can be derived from $\dot{Q} = \mathrm{Tr}\left\{H_0 \mathcal{L}[\rho]\right\} = \dot{Q}_{\rm{r}} + \dot{Q}_{\rm{n}}$, where $\mathcal{L}[\rho]$ is the {dissipative part of the Liouvillian} for the interaction with both baths [Supplementary Material S2],  and $\dot{Q}_{\rm{i}}=  \hbar\omega_{01}(\Gama{i}n_{\rm{i}}\ex{\sigma_-\sigma_+}-\Gama{i}(n_{\rm{i}}+1)\ex{\sigma_+\sigma_-}$) $\rm{i}\in\{\rm{n},\,\rm{r}\}$. In particular, we find from Eq.~(\ref{outp}) that 
$\dot{Q}_{\rm{r}}=P_{\rm{loss}}-\dot{W}$ [blue in Fig.~\ref{powerloss}(c)], {where the negative values of $\dot{Q}_{\rm{r}}$ means that the waveguide is heated up due to the qubit emission}. Interestingly, when the drive is off ($\dot{W}=0$), and in the steady-state (the change of the internal energy of the qubit $\dot{U}=\mathrm{Tr}\left\{H_0\partial_{t}\rho(t)\right\}=0$), the first law of thermodynamics $\dot{Q}_{\rm{r}}+\dot{Q}_{\rm{n}}+\dot{W}=\dot{U}$ implies that we have $\dot{Q}_{\rm{n}}=-\dot{Q}_{\rm{r}}$, indicating the heat exchange between these two baths, consistent with what we observed.

{\bf{Noise origin.}} {In a separate cooldown, we observe that improving the isolation between the sample and the first amplifier in our chain [see Methods] leads to a strong reduction in the nonequilibrium heat flow, and a reduced thermal population of the qubit [red stars in Fig.~\ref{powerloss}(a)]. This observation suggests that} the temperature of the non-radiative environment has been reduced, with $\Gama{n}/2\pi=\unit[53\pm4]{kHz}$, and that the two baths are in thermal equilibrium without external drive. Therefore, we have a reduced thermal population of our qubit: $\rho_{11}^{\rm{th}}\approx \unit[0.4]{\%}$.  {
It is possible that noise from the high-mobility electron transistor amplifier (HEMT) at the 4K stage induces quasiparticles~\cite{wenner2013excittion, serniak2018hot} or active TLSs~\cite{de2020two, oliver2013materials}.

Quasiparticles may in principle be generated from stray thermal noise propagating to the sample from the output line. These ``hot'' quasiparticles with  energy higher than $\Delta_{\rm{g}}+\hbar\omega_{01}$ could excite the qubit~\cite{wenner2013excittion, serniak2018hot, jin2015}, where $\Delta_{\rm{g}}$ is the superconducting gap. However, from the thermal population of our qubit $\rho_{11}^{\rm{th}}\approx \unit[2.86]{\%}$, we can infer that the quasiparticle-induced nonradiative decay rate should be $\Gamma_{\rm{qp}}/2\pi\approx \unit[84]{kHz}>\Gama{n}$ (see Methods section), in conflict with our observations. Moreover, if the excitation is dominated by the quasiparticles, the value of $\Gama{n}$ should be decreased after we {decrease the temperature of} the hot bath which is not the case in our experiment. Conversely, we do not see a substantial increase either, as might be expected for a purely TLS bath (see Methods).  Thus, we believe our hot bath is primarily from excited two-level systems, but quasiparticles may still play a small role. However, precisely reconstructing the properties and origin of this thermal noise requires further investigation. We also point out that our model makes it possible to study the energy exchange between ``hot'' quasiparticles and a cold waveguide, if the waveguide is cold enough [See Methods].

\section{Discussion}

In summary, with the system parameters shown in Table~\ref{tab:1}, we constructed a two-bath model to explain the heat transport via a superconducting qubit. Our study indicates that the qubit excited-state population is increased up to \unit[2.9]{\%} due to interaction with a hot environment, probably from excited TLSs, with an equivalent temperature of about $\unit[131]{mK}$. This temperature can be reduced by adding more isolation between the sample and the output line, leading to the thermal population of our qubit being reduced {by almost one order of magnitude}. {Our results point at the necessity of providing sufficient isolation between the sample and the HEMT amplifier to mitigate residual thermal occupation of qubits, which adversely affects state preparation in quantum computing applications~\cite{mcewen2021removing, Egger2018}.}

{Moreover, as presented in the master equation, the noise will increase the intrinsic qubit relaxation rate by $\Gama{+}=n_{\rm{n}}\Gama{n}+n_{\rm{r}}\Gama{r}$. Thus, $\Gama{+}$ is reduced from about \unit[9]{kHz} with two isolators to \unit[1]{kHz} with four isolators, which is about \unit[15]{\%} of the nonradiative decay rate. Such a large improvement will be important in circuit QED, where generally the radiative decay $\Gama{r}$ is negligible and $\Gama{n}$ dominates the qubit decay.}

We also showed that, under strong drive, we can use the waveguide to measure work and heat rates. While our quantum system is operating in a regime where the drive only increases the heat rate to both environments, having direct access to the work, from measuring the coherent output, and the heat, from the power loss, is a powerful tool for the future study of regimes where useful tasks like cooling or work extraction can be performed. 

{\bf{Noise spectrometer.}} Based on the model we developed here, we propose that our study also enables the construction of a practical noise spectrometer. 
In detail, we could engineer two waveguide channels coupled to the qubit, similar to the setup in Refs.~\cite{peng2016tuneable, zhou2020tunable}, but with equally strong coupling on both sides with $\Gama{1}=2\Gama{r}$ ($\Gama{r}\gg\Gama{n}$). One channel, acting as a probe, could be used for inputting noise while the other, acting as a detector, is for measuring the noise PSD. In this setup, the noise can be very well isolated from the detection channel since the direct capacitive coupling between the probe and detection channels can be designed to be extremely small~\cite{peng2016tuneable, zhou2020tunable}.

In this scenario, the thermal spectrum density in Eq.~(\ref{thermalspectrum}) becomes

\beq
S_{\rm{th}}(\omega) =\hbar\omega_{01}\frac{\Gama{r}}{2\pi}\frac{\Gama{2}\Delta n}{\delta\omega_{01}^2 + \Gama{2}^2},
\label{thermalspectrumthermometer}
\eeq
where $\Delta n=n_{\rm{th}}-n_{\rm{r}}$, with $n_{\rm{th}}$ ($n_{\rm{r}}$), the temperature of the probe (detection) channel. We notice that Eq.~(\ref{swon}) is still valid. Thus, we can evaluate the value of $\Delta n$ at the qubit frequency according to $\Delta n=S_{\rm{th}}(\omega)/2S_{\rm{on}}(\omega)$ if $S_{\rm{th}}(\omega)$ and $S_{\rm{on}}(\omega)$ are measured. Furthermore, by taking the integral of the spectra, we have the powers $P_{\rm{th}}=\hbar\omega_{01}\Gama{r}\Delta n/2$ and $P_{\rm{on}}=\hbar\omega_{01}\Gama{r}/4$ for $S_{\rm{th}}$ and $S_{\rm{on}}$, respectively. Therefore, we obtain $\Delta n=P_{\rm{th}}/2P_{\rm{on}}$.

Compared to measuring the whole spectra, the second method will reduce the measurement time by a factor  proportional to the number of data points in the spectra. However, by observing the symmetry of the spectra, we can check whether the bandwidth of the noise is larger than $\Gama{r}$ or not, which can be possible when a qubit strongly couples to some resonant TLSs~\cite{you2021positive}. Note that it is not necessary to calibrate the system gain with these two measurements. Moreover, since the method is insensitive to $\Gama{\phi}$, {this insensitivity makes the method robust against flux noise, which introduces excess dephasing when the qubit frequency is tuned away from the sweet spot. As a result, the spectrometer can be reliably operated over a broad frequency range}. Combined with the reflection coefficient, we will have $r=n_{\rm{th}}+n_{\rm{r}}$, when $\Gama{\phi}\ll\Gama{r}$. Thus, we can obtain $n_{\rm{th}}$ and $n_{\rm{r}}$, separately.

Compared to thermometers based on tunneling junctions~\cite{spietz2003primary, wang2018fast}, our proposal does not require additional calibration and enables the possibility to obtain the noise spectrum by sweeping the qubit frequency. Circuit-QED radiometers~\cite{wang2021radiometry, xu2020radiative} have also been developed, where the bandwidth is limited by the cavity. Very recently, a thermometer based on the reflection coefficient of a qubit in front of a mirror was demonstrated~\cite{scigliuzzo2020primary}, where it has a single bosonic bath in the waveguide, namely, $\Gamma_{\rm{r}}\neq0$ ($n_{\rm{r}}\neq0$) with $\Gamma_{\rm{n}}=0$ ($n_{\rm{n}}=0$). {Compared to Ref.~\cite{scigliuzzo2020primary}, the thermometry method suggested here, based on PSD measurements, can tolerate a higher noise background in the probe waveguide, and, unlike Ref.~\cite{scigliuzzo2020primary}, is insensitive to the pure dephasing rate.} Finally, separating the detection and probe channels can make it easier to connect our thermometer chip directly to other hybrid quantum systems~\cite{albanese2020radiative, mirhosseini2020superconducting, Fink2020converter, han2020cavity}.

\section{Methods}

{\bf{Measurement of PSD}}. To obtain the PSD, we measure the output signal of the qubit emission into the waveguide as a voltage in the time domain, normalized to the system gain based on the Mollow-triplet spectrum~\cite{astafiev2010resonance}, and then calculate the PSD according to the Welch method~\cite{welch1967use}, where the background noise is subtracted by the reference found by tuning the qubit frequency away using the external magnetic flux. Experimentally, in order to remove the background and the gain drift, we first measure the PSD for \unit[1]{s} with the drive either on or off, as required. After that, we tune the qubit away by changing the external flux and repeat the measurement to obtain the background reference. Finally, we just take the difference between these two measurements to obtain either $S_{\rm{on}}$ or $S_{\rm{off}}$, if the drive was on or off, respectively.

{\bf{Quasiparticles}}. In addition to  the hot bath we discussed, nonequilibrium quasiparticles could also excite and decay the qubit with rates $\Gama{\uparrow}$ and $\Gama{\downarrow}$, respectively. Therefore, we add two additional dissipators ($\Gama{\downarrow}\mathcal{D}[\sigma_-]\rho $ and $\Gama{\uparrow}\mathcal{D}[\sigma_+]\rho $) into the master equation used for our model [Supplementary Material S2]. By solving the master equation without external drive, we obtain the lost power, which contains the effects from both the hot bath and quasiparticles as
\beq
\frac{P_{\rm{loss}}}{\hbar\omega_{01}} = \frac{\Gama{r}\Gama{n}(n_{\rm{r}}- n_{\rm{n}})-\Gama{r}[(n_{\rm{r}}+1)\Gama{\uparrow}-n_{\rm{r}}\Gama{\downarrow}]}{\Gama{1}+\Gama{\uparrow}+\Gama{\downarrow}}.
\eeq
When $\Gama{n}=0$ and $n_{\rm{r}}=0$, the qubit population due to the nonequilibrium quasiparticles is $\rho_{11}^{\rm{qp}}=\frac{\Gama{\uparrow}}{\Gama{\uparrow}+\Gama{\downarrow}}$.
Experimentally, we can increase the capacitive coupling between the waveguide and the qubit in order to make the nonradiative decay negligible ($\Gama{r}\gg\Gama{n}$). By suppressing the noise from the output line, as we did in the main text, we have $n_{\rm{r}}\approx n_{\rm{n}}$. In these conditions, the lost power is mainly from the quasiparticles. Depending on the value of the waveguide thermal-photon occupation, we have three regimes: if $n_{\rm{r}}>\Gama{\uparrow}/(\Gama{\downarrow}-\Gama{\uparrow})$, $P_{\rm{loss}}>0$, the qubit is thermally excited and prefers to be dissipated due to the quasiparticle tunneling through the Josephson junction; if $n_{\rm{r}}<\Gama{\uparrow}/(\Gama{\downarrow}-\Gama{\uparrow})$, $P_{\rm{loss}}<0$, the quasiparticles excite the qubit, and then the qubit emits a photon into the waveguide; if $n_{\rm{r}}=\Gama{\uparrow}/(\Gama{\downarrow}-\Gama{\uparrow})$, $P_{\rm{loss}}=0$, we can consider the effects on the qubit from the quasiparticles and the thermal photons in the waveguide are the same.

The quasiparticle-induced excited-state population can be approximately written as~\cite{wenner2013excittion}
\beq
\rho_{11}^{\rm{qp}}\simeq 2.17\frac{n_{\rm{qp}}}{n_{\rm{cp}}}\left(\frac{\Delta_{\rm{g}}}{\hbar\omega_{01}}\right)^{3.65}, \label{quasi}
\eeq
in which $n_{\rm{qp}}$ ($n_{\rm{cp}}$) is the density of all quasiparticles (Cooper pairs) and $\Delta_{\rm{g}}$ is the superconducting energy gap. Combining this with the quasiparticle-induced decay rate for a transmon qubit~\cite{wenner2013excittion, Catelani2011quasiparticle}
\beq
\Gama{\downarrow}\simeq \frac{\sqrt{2}}{R_{N}C}\frac{n_{\rm{qp}}}{n_{\rm{cp}}}\left(\frac{\Delta_{\rm{g}}}{\hbar\omega_{01}}\right)^{1.5}, \label{quasi2}
\eeq
we have
\beq
\Gama{\downarrow}\simeq \frac{\sqrt{2}}{2.17R_{N}C}\left(\frac{\Delta_{\rm{g}}}{\hbar\omega_{01}}\right)^{-2.15}\cdot\rho_{11}^{\rm{qp}}, \label{quasi3}
\eeq
where the normal resistance of our SQUID is $R_{N}\approx\unit[6.3]{k\Omega}$, the total capacitance of the qubit is $C=\unit[78]{fF}$  and $\Delta_{\rm{g}}=\unit[170]{\mu eV}$ for aluminium. By assuming that our thermal population is solely from the quasiparticles, i.e., $\rho_{11}^{\rm{qp}}=\rho_{11}^{\rm{th}}\approx \unit[2.86]{\%}$, we have $\Gama{\downarrow}/2\pi\approx \unit[87]{kHz}$. Therefore, we have $\Gama{\uparrow}\approx \rho_{11}^{\rm{qp}}\Gama{\downarrow}\approx\unit[2\pi\times 3]{kHz}$. Thus, the quasiparticle-induced nonradiative decay rate would be $\Gama{qp}=\Gama{\downarrow}-\Gama{\uparrow}\approx2\pi\times \unit[84]{kHz}>\Gama{n}$, see discussion in the main text.
\begin{figure}[tbph]
\centering
\includegraphics[width=1\linewidth]{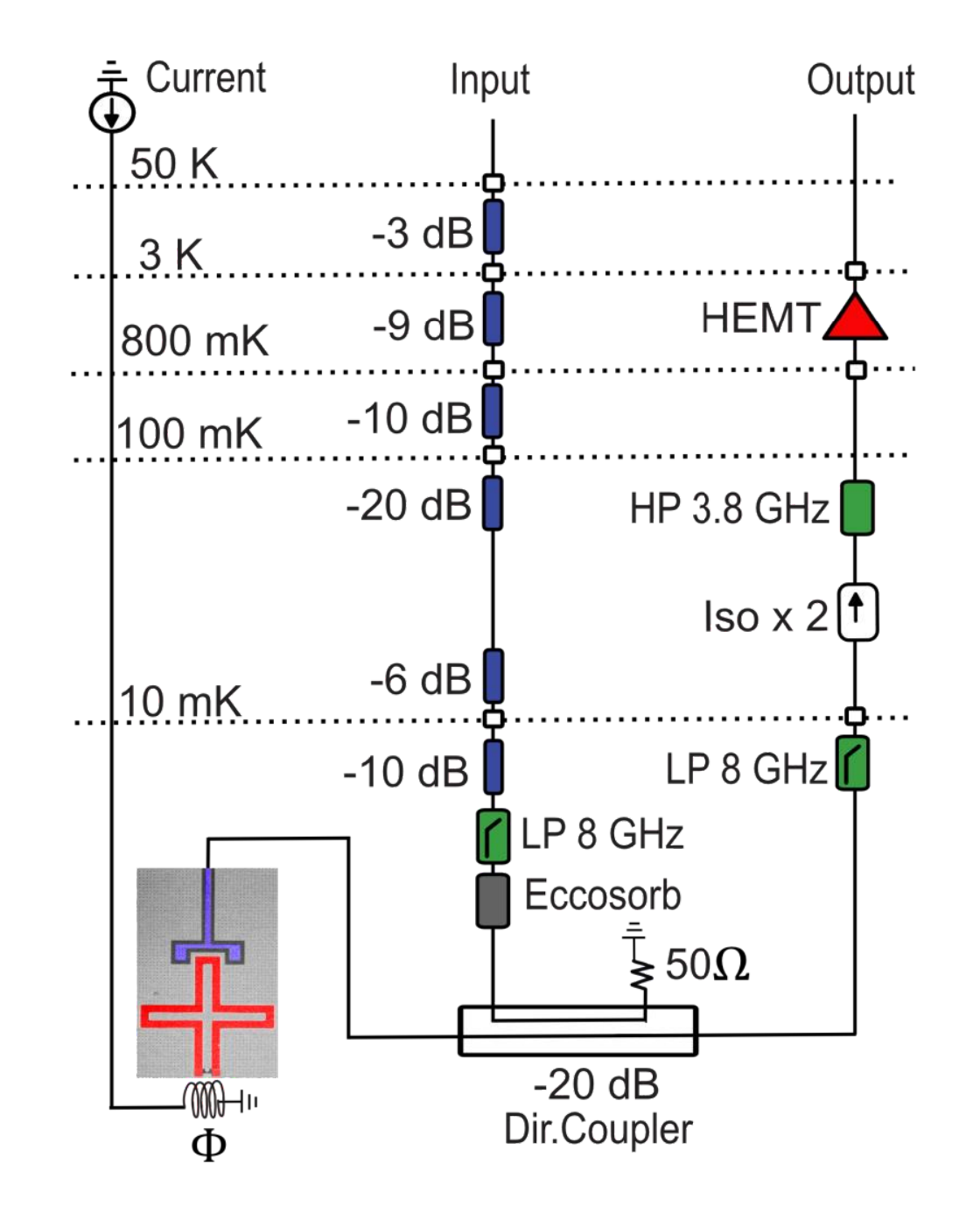}
\caption{\textsf{{\bf{\textsf{Experimental setup}}.}
Setup inside the dilution refrigerator. The signal from the input port is attenuated and fed into the waveguide through a low-pass filter (LP), Eccosorb, and a -20\unit{dB} directional coupler. After interacting with the transmon,  the signal from the sample goes through the directional coupler again, passes another LP, two isolators (Iso), and a high-pass filter with amplification from a HEMT amplifier. The qubit frequency can be changed by sending a current to the magnetic flux coil to generate an external magnetic flux $\Phi$ through the SQUID.
}
}
\label{setup}
\end{figure}

{Finally, as discussed in the main text, we do not observe any considerable decrease in $\Gama{n}$ as the temperature is lowered (by the addition of isolators). Conversely, we also notice that the value of $\Gama{n}$ does not observably increase. If we assume that $\Gama{n}$ is solely from TLSs, then, when we decrease the bath temperature from \unit[131]{mK} to \unit[50]{mK}, the value of $\Gama{n}$ is expected to increase by about \unit[10]{kHz} according to the relationship $\Gama{n,T}=\Gama{n,0}\tanh(\hbar\omega_{01}/k_{\rm{B}}T)$~\cite{muller2019towards} (see also Supplementary Information), where $\Gama{n,0}$ is the nonradiative decay rate due to TLSs at zero temperature. It may be that the high-temperature noise from the output may also generate some quasiparticles. Assuming that the noise temperature is the same as that of the TLS bath, namely, \unit[131]{mK}, the corresponding quasiparticle induced decay rate is about \unit[8]{kHz}, as given by $\Gamma_{\rm{qp}}=\frac{\omega_{01}}{\pi}\sqrt{\frac{2\Delta_{\rm{g}}}{\hbar\omega_{01}}}x_{\rm{qp}}$ with the normalized quasiparticle density $x_{\rm{qp}}=\frac{\sqrt{2\pi\Delta_{\rm{g}}k_{\rm{B}}T}}{\Delta_{\rm{g}}}e^{-\Delta_{\rm{g}}/k_{\rm{B}}T}$ \cite{Catelani2011quasiparticle,barends2011minimizing}, and said rate would decrease as the noise temperature is lowered. In total, it is possible that we do not see a change in $\Gama{n}$ after suppressing the noise because these two effects (TLS rate increasing, quasiparticle rate decreasing) mitigate each other.}

{\bf{Measurement setup}}.
{Figure~\ref{setup} shows our measurement setup, where a transmon qubit is weakly coupled to a 1D semi-infinite transmission line. To measure the reflection coefficient, a vector network analyzer (VNA, not shown) generates coherent continuous microwaves which are sent into the input line. This signal is attenuated, pass through a $20~{\rm{dB}}$ directional coupler, and then reaches the qubit. After thes interaction with the qubit, the electrical field is reflected back, and passes filters, isolators and a HEMT amplifier. Finally, it goes back to the VNA to obtain the reflection coefficient with some room-temperature amplifiers.}

{To measure the power spectral density (PSD) from the thermal noise, turning off the VNA, we use a digitizer to measure the voltage from the output port in the time domain where the sampling rate is 3~MHz and 4~MHz for Fig.\,3(a) and (b), respectively. Such rates are large enough to capture the signal around the qubit in the frequency domain. After that, as discussed above about the measurement of the PSD, we take the Fourier transform to obtain the PSD by using the Welch method\,\cite{welch1967use}. The PSD with strong drive on the qubit is obtained in the same way.}

\section*{Contributions}
Y.L.~planned and performed the measurements. A.B.~and Y.L.~designed the sample, A.B.~fabricated the device. N.L.~and Y.L.~developed the model.
N.L., Y.L., K.F., and A.F.K.~performed the theoretical analysis. Y.L.~and N.L.~wrote the manuscript with help from all authors. P.D.~supervised this work.

\acknowledgements
We acknowledge fruitful discussions with Shahnawaz Ahmed and Dr. Jonathan Burnett. {We also thank Dr.~Niklas Wadefalk and Dr.~Sumedh Mahashabde for useful discussions about the HEMT amplifiers.} Numerical modelling was performed using the QuTiP library~\cite{qutip,qutip2}. N.L.~acknowledges partial support from JST PRESTO through Grant No.~JPMJPR18GC. F.N.~is supported in part by: NTT Research, Japan Science and Technology Agency (JST) (via the Q-LEAP program, Moonshot R\&D Grant No.~JPMJMS2061, and the CREST Grant No.~JPMJCR1676), Japan Society for the Promotion of Science (JSPS) (via the KAKENHI Grant No.~JP20H00134 and the JSPS-RFBR Grant No.~JPJSBP120194828), Army Research Office (ARO) (Grant No.~W911NF-18-1-0358), Asian Office of Aerospace Research and Development (AOARD) (via Grant No.~FA2386-20-1-4069). F.N.~and N.L.~acknowledge the Foundational Questions Institute Fund (FQXi) via Grant No.~FQXi-IAF19-06. We~acknowledges the use of the Nanofabrication
Laboratory (NFL) at Chalmers. Y.L., A.F.K., S.G., and P.D.~were supported by the Knut and  Alice  Wallenberg  Foundation  through  the  Wallenberg Center for Quantum Technology (WACQT). Y.L.~and P.D.~are supported by the Swedish Research Council [VR R\aa{}dsprof (5920793)].

\bibstyle{apsrev4-1}
%

\widetext
\clearpage
\begin{center}
\textbf{\large Supplementary Material for "Nonequilibrium heat transport and work with a single artificial atom coupled to a waveguide: emission without external driving"}\\
\vspace{4mm}

Yong Lu$\text{n}^{1,*}$, Neill Lambert$\text{n}^{2,*}$, Anton Frisk Kockum$\text{}^{1}$, Ken Funo$\text{}^{2}$, Andreas Bengtsson$\text{}^{1}$, Simone Gasparinetti$\text{}^{1}$, Franco Nori$\text{n}^{2,3}$, Per Delsing$\text{}^{1}$\\
\vspace{2mm}

{\it {\small $^{1}$Microtechnology and Nanoscience, MC2, Chalmers University of Technology, SE-412 96 G\"oteborg, Sweden}} \\
{\it {\small $^{2}$Theoretical Quantum Physics Laboratory, RIKEN Cluster for Pioneering Research, Wako-shi, Saitama 351-0198, Japan}}\\
{\it {\small $^{3}$Department of Physics, The University of Michigan, Ann Arbor, 48109-1040 Michigan, USA}}
\end{center}

\setcounter{tocdepth}{0}

\renewcommand{\thefigure}{S\arabic{figure}}
\renewcommand{\thesection}{S\arabic{section}}
\renewcommand{\theequation}{S\arabic{equation}}

\onecolumngrid

\setcounter{section}{0}
\setcounter{equation}{0}

\setcounter{figure}{0}

\section{Master equation}
The dynamics and steady state of the qubit in contact with two bosonic heat baths can be found, in the weak-coupling and Markovian approximations, by solving the equation of motion

\beq
\frac{\partial}{\partial t} \rho_S(t) &=& -\frac{i}{\hbar}[H_{\rm{q}},\rho(t)] + \mathcal{L}[\rho(t)], \label{2levME}
\eeq
where
\beq
\mathcal{L}[\rho(t)] &=& \frac{\Gama{r}}{2} \left(n_{\rm{r}}+1\right)\mathcal{D}[\sigma_-]\rho+\frac{\Gama{r}}{2} n_{\rm{r}}\mathcal{D}[\sigma_+]\rho+ \frac{\Gama{n}}{2} \left(n_{\rm{n}}+1\right)\mathcal{D}[\sigma_-]\rho+\frac{\Gama{n}}{2} n_{\rm{n}}\mathcal{D}[\sigma_+]\rho + \frac{\Gama{\phi}}{4}\mathcal{D}[\sigma_z]\rho,\label{lind}
\eeq
where the Lindblad operator is $\mathcal{D}[\sigma_i]\rho(t) = 2\sigma_i \rho(t)\sigma_i^{\dagger} - \{\sigma_i^{\dagger}\sigma_i,\rho(t)\}$, and the thermal occupation of the bosonic baths are given by
$n_i = [\exp(\hbar\omega_{01}/k_{\mathrm{B}}T_i) -1]^{-1}$, where $k_{\mathrm{B}}$ is the Boltzmann constant.

The qubit Hamiltonian  is
\beq
\frac{H_{\rm{q}}}{\hbar} = -\frac{\Delta}{2}\sigma_z + \frac{\Omega}{2}\sigma_x,
\eeq
 where the detuning $\Delta = \omega_{\rm{p}}-\omega_{01}$ is the energy difference between the drive at frequency $\omega_{\rm{p}}$ and the bare qubit frequency $\omega_{01}$.

In the above, we explicitly assume that the qubit only has two levels, and that the radiative (transmission line) and non-radiative baths are bosonic, and obey the Born-Markov secular (BMS) approximation.  The influence of a third level is described below.

The second assumption, the bosonicity of the non-radiative bath, can be replaced by assuming a TLS bath.  This simply changes the temperature dependence of the Lindblad operators for that bath, such that
\bea
\frac{\Gama{n}}{2}\left(n_{\rm{n}}+1\right)\mathcal{D}[\sigma_-]\rightarrow \frac{\Gama{n}}{2}\left(\frac{1+n_{\rm{n}}}{1+2 n_{\rm{n}}}\right)\mathcal{D}[\sigma_-], \,\,\,\,\,\,\frac{\Gama{n}}{2}\left(n_{\rm{n}}\right)\mathcal{D}[\sigma_+]\rightarrow \frac{\Gama{n}}{2}\left(\frac{n_{\rm{n}}}{1+2 n_{\rm{n}}}\right)\mathcal{D}[\sigma_+].
\eea

Ultimately, this implies some ambiguity in the temperature ascertained from the height of the thermal peak in the spectrum calculation. For simplicity we use that obtained from the bosonic assumption, but note that a TLS bath requires a larger temperature to produce the equivalent thermal peak in the spectrum, since $k_{\mathrm{B}}T_n = \hbar \omega_{01}/[\ln(1\pm n_{\rm{n}})-\ln(n_{\rm{n}})]$, where $\pm = +$ for a bosonic bath, and $\pm =-$ for a TLS bath. For example, for the $\Delta n =0.135$ observed in the main text we would find, assuming $n_{\rm{r}} = 0$, $k_{\mathrm{B}}T_n/\hbar\omega_{01}=0.47$ for a bosonic bath and $0.54$ for a TLS one.

It is convenient to use the Heisenberg equations of motion generated by  \eqref{2levME}  for the operators $s_1(t)=\ex{\sigma_-(t)}$, $s_1^*(t)=\ex{\sigma_+(t)}$, and $s_2(t) = \ex{\sigma_+(t)\sigma_-(t)}$:

\beq
 \frac{d}{dt} \begin{pmatrix}
  s_1     \\
  s_1^{*} \\
  s_2
\end{pmatrix} =
M  \begin{pmatrix}
  s_1     \\
  s_1^{*} \\
  s_2
\end{pmatrix} + B, \label{heom2}
\eeq
where
\beq
M =  \begin{pmatrix}
   i\Delta - \Gama{2}&0 &i\Omega \\
   0&-i\Delta - \Gama{2} & -i\Omega^* \\
   i\Omega^*/2 & -i\Omega/2 &-\Gama{1}
\end{pmatrix}
\eeq
and
\beq
B = \begin{pmatrix}
  -i\Omega/2     \\
  i\Omega^*/2 \\
  \Gama{+}
\end{pmatrix}
\eeq
and where, as in the main text, $\Gama{+}=n_{\rm{n}}\Gama{n}+n_{\rm{r}}\Gama{r}$, $\Gama{1}=(1+2n_{\rm{n}})\Gama{n}+(1+2n_{\rm{r}})\Gama{r}$, and $\Gama{2}=\Gama{\phi}+\Gama{1}/2$. For the steady state $t\rightarrow \infty $ this gives
\beq
\ex{\sigma_-(t\rightarrow \infty)} = \frac{\Omega (\Gama{1} - 2 \Gama{+})\left(\Delta - i\Gama{2}\right)}{2\Omega^2\Gama{2} + 2 (\Delta^2 + \Gama{2}^2)\Gama{1}} \label{sm}
\eeq

\beq
\ex{\sigma_+\sigma_-(t\rightarrow \infty)} = \frac{|\Omega|^2 \Gama{2} + 2 \Gama{+}(\Delta^2 + \Gama{2}^2)}{2|\Omega|^2\Gama{2} + 2 (\Delta^2 + \Gama{2}^2)\Gama{1}} \label{spsm}
\eeq
\beq
\ex{\sigma_-\sigma_+(t\rightarrow \infty)} = \frac{|\Omega|^2 \Gama{2} + 2 \Gama{-}(\Delta^2 + \Gama{2}^2)}{2|\Omega|^2\Gama{2} + 2 (\Delta^2 + \Gama{2}^2)\Gama{1}}, \label{smsp}
\eeq
where in the last equation $\Gama{-}=(n_{\rm{n}}+1)\Gama{n}+(n_{\rm{r}}+1)\Gama{r}$.

\section{Reflectivity}

To calculate the reflectivity, we use \eqref{sm} with
\beq
r = \frac{b_{\mathrm{out}}}{b_{\mathrm{in}}} = 1 - \frac{i2\Gama{r}}{\Omega} \ex{\sigma_-(t\rightarrow \infty)},
\eeq
where assuming weak drive $\Omega \ll \Gamma_1$ gives
\beq
r=1 - i\Gama{r}\frac{(1-2\Gama{+}/\Gama{1} )}{\Delt+i\Gama{2}}.
\label{reflection2}
\eeq

\section{Power output}
Recalling from the main text, we employ input-output theory for the output of the transmission line $b_{\mathrm{out}}(t) =   b_{\mathrm{in}}(t) - i\sqrt{\Gama{r}} \sigma_-(t) $, where $b_{\mathrm{in}}(t) = f_{\mathrm{in}}(t) + \frac{\Omega}{2\sqrt{\Gama{1}}}$ includes the thermal noise $f_{\mathrm{in}}(t)$ and coherent drive $\frac{\Omega}{2\sqrt{\Gama{1}}}$. To obtain the correct expression for the output power (and the power spectral density) we need to calculate correlations between the input thermal field in the waveguide and the qubit of the form $\ex{\sigma_+(t) f_{\mathrm{in}}(t')}$. We can can evaluate these using the result from Ref.~\cite{Gardiner04},  where, assuming the effect of the waveguide on the dynamics of the system obeys the standard  BMS approximation, it is possible to show that for $t<t'$, the input field has not yet interacted with the qubit, so $\ex{\sigma_+(t) f_{\mathrm{in}}(t')} =0$.  For $t=t'$ and $t>t'$, the thermal input can be correlated with the qubit, and the following holds:

\beq
\ex{\sigma_+(t)f_{\mathrm{in}}(t')}  = -i \sqrt{\Gamma_{\rm{r}}}n_{\rm{r}} \Theta(t-t') \ex{\left[\sigma_+(t), \sigma_-(t')\right]}
\eeq
and
\beq
\ex{f^{\dagger}_{\mathrm{in}}(t) \sigma_-(t')} = i \sqrt{\Gamma_{\rm{r}}}n_{\rm{r}} \Theta(t'-t) \ex{\left[\sigma_+(t),\sigma_-(t') \right]} .
\eeq

Splitting the input field into thermal terms and coherent terms, one can evaluate the output field intensity correlator as

\beq\label{outp3}
\langle  b_{\mathrm{out}}^{\dagger}(t) b_{\mathrm{out}}(t')\rangle &=& \Gama{r}\left(n_{\rm{r}}+1\right)\langle \sigma_+(t) \sigma_-(t')\rangle
- \Gama{r} n_{\rm{r}}
\langle  \sigma_-(t') \sigma_+(t)\rangle + \langle f_{\mathrm{in}}^{\dagger}(t)f_{\mathrm{in}}(t')\rangle \nonumber\\ &-& \frac{i\Omega^*}{2}\langle \sigma_-(t')\rangle + \frac{i\Omega}{2} \langle\sigma_+(t)\rangle +  \frac{|\Omega|^2}{4\Gama{r}}.
\eeq
With this we can also evaluate the equal-time terms for the power output, but it is more instructive to show the evaluation more explicitly, following the steps outlined in Ref.\,\cite{CarmichaelI} modified to accommodate a qubit instead of cavity.

 As described in the main text, we start with the full system Hamiltonian for the qubit and the two environments,
\beq
H_{\rm{sys}} = H_{\rm{q}} + H_{\rm{r}} + H_{\rm{n}}.
\eeq
The radiative ($H_{\rm{r}}$) and non-radiative bath Hamiltonians ($H_{\rm{n}}$) are
\beq
\frac{H_{{i}}}{\hbar}=\sum_{{k}} \omega_{{k,i}} a_{{k,i}}^{\dagger}a_{{k,i}} + \sum_k g_{{k,i}} \left(\sigma_- a_{{k,i}}^{\dagger} + \sigma_+a_{{k,i}} \right),
\eeq
and include the interaction with the qubit.

 First we absorb the non-radiative bath into the qubit Hamiltonian,
\beq
H'_{\rm{q}} = H_{\rm{q}} + H_{\rm{n}}.
\eeq
This gives us Langevin equations  for the system operators coupled to the waveguide (setting $\hbar =1$ in the following steps for simplicity),
\beq
\dot{\sigma}_- = -i[\sigma_-,H_{\rm{sys}}] = -i[\sigma_-,H'_{\rm{q}}] -i[\sigma_-,\sigma_+] \sum_{\rm{k}} g_{{k,r}} a_{{k,r}}.\label{heom}
\eeq

We then define the spectral density for the transmission line as $J(\omega)_{\mathrm{r}}=\pi\sum_{{k}}|g_{{k,r}}|^2 \delta(\omega-\omega_{{k,r}})$, insert the definition of the integral of the transmission-line modes into \eqref{heom}, and
make the standard Markovian approximation $J(\omega)_{\mathrm{r}} = \Gamma_{\mathrm{r}}$.  Evaluating the result gives
\beq
\dot{\sigma}_- = -i[\sigma_-,H'_{\rm{sys}}] -\frac{\Gamma_{\rm{r}}}{2}\sigma_- +i\sigma_z \sqrt{\Gamma_{\rm{r}}}b_{\mathrm{in}}(t).
\eeq

To obtain correlation functions between the thermal input and system operators, we omit the coherent contribution to $b_{\mathrm{in}}(t) = f_{\mathrm{in}}(t) + \frac{\Omega}{2\sqrt{\Gama{1}}}$ and multiply the equation of motion from the left first with $\sigma_z$ and then with the required system operator for the correlation function we wish to evaluate, say $\sigma_+$. Rearranging gives the form
\beq
i \sqrt{\Gamma_r} \ex{\sigma_+f_{\mathrm{in}}} &=&  -\ex{\sigma_+\dot{\sigma}_-}-\frac{\Gamma_{\rm{r}}}{2}\ex{\sigma_+\sigma_-}-i \ex{\sigma_+[\sigma_-,H'_{\rm{q}}]} \label{eomcorr}
\eeq

To evaluate the term $\ex{\sigma_+\dot{\sigma}_-}$ we use the formula

\beq
2\ex{O_1 O_2} = \frac{d}{dt}\ex{O_1O_2} + \rm{Tr}\left\{O_2\mathcal{L}_r[\rho O_1] - O_1\mathcal{L}_r[O_2\rho]\right\},
\eeq
where $\rho$ is the system density matrix after tracing out the waveguide, and $\mathcal{L}_r$ is the Liouvillian describing the evolution of the system in contact with the waveguide,
\beq
\mathcal{L}_r[\rho] = -i[H'_{\rm{sys}},\rho] + \frac{\Gama{r}}{2} \left(n_{\rm{r}}+1\right)\mathcal{D}[\sigma_-]\rho+\frac{\Gama{r}}{2} n_{\rm{r}}\mathcal{D}[\sigma_+]\rho.
\eeq
Evaluating this gives
\beq
\ex{\sigma_+\dot{\sigma}_-} = -i \ex{\sigma_+[\sigma_-,H'_{\rm{sys}}]} -\frac{\Gama{r}}{2} \ex{\sigma_+\sigma_-} - \frac{\Gama{r}n_{\rm{r}}}{2}\ex{[\sigma_+,\sigma_-]}
\eeq
Combining this with \eqref{eomcorr} we find
\beq
i \sqrt{\Gamma_r} \ex{\sigma_+f_{\mathrm{in}}} &=&\frac{\Gama{r}n_{\rm{r}}}{2}\ex{[\sigma_+,\sigma_-]}\label{p1}.
\eeq
Performing the same steps for the other system-thermal noise correlator, and inserting into the output-power formula, we find, as expected
\beq\label{outp2}
\langle  b_{\mathrm{out}}^{\dagger} b_{\mathrm{out}}\rangle &=& \Gama{r}\left(n_{\rm{r}}+1\right)\langle \sigma_+ \sigma_-\rangle- \Gama{r} n_{\rm{r}}
\langle  \sigma_- \sigma_+\rangle + \langle f_{\mathrm{in}}^{\dagger}f_{\mathrm{in}}\rangle \nonumber\\ &-& \frac{i\Omega^*}{2}\langle \sigma_-\rangle + \frac{i\Omega}{2} \langle\sigma_+\rangle +  \frac{|\Omega|^2}{4\Gama{r}}.
\eeq

One can use this to show that the power loss $P_{{\rm{loss}}}=\hbar\omega_{01}\left(\langle  b_{\mathrm{in}}^{\dagger} b_{\mathrm{in}}\rangle- \langle  b_{\mathrm{out}}^{\dagger} b_{\mathrm{out}}\rangle \right)$  under zero drive ($\Omega =0$) is given by
\beq
P_{\rm{loss}} = \frac{\hbar\omega_{01}\Gama{r}\Gama{n}(n_{\rm{r}}- n_{\rm{n}})}{\Gama{1}}.
\eeq

\section{Power spectral density}

As described in the main text, to evaluate the power spectral density
\beq
S(\omega) =\frac{\hbar\omega_{01}}{2\pi} \int_{-\infty}^{\infty} dt e^{-i\omega t} \langle  b_{\mathrm{out}}^{\dagger}(t) b_{\mathrm{out}}(t')\rangle_{(t' \rightarrow \infty)} \label{sw}
\eeq
we need to evaluate the correlation functions contained in \eqref{outp3} for the system operators. In some simple cases, this can be done by hand. For example, in the zero-drive limit, one can easily show that
\beq
\ex{\sigma_+(t)\sigma_-(t')}_{t'\rightarrow \infty} = e^{-\Gama{2}|t|}e^{ - i\Delta t} \ex{\sigma_+\sigma_-(t')}\nonumber
\eeq
and
\beq
\ex{\sigma_-(t')\sigma_+(t)}_{t'\rightarrow \infty} = e^{-\Gama{2}|t|}e^{ - i\Delta t} \ex{\sigma_-\sigma_+(t')},\nonumber
\eeq
where $\ex{\sigma_+\sigma_-(t')}_{t'\rightarrow \infty} = \frac{\Gama{+}}{\Gama{1}}$ and $\ex{\sigma_-\sigma_+(t')}_{t' \rightarrow \infty} =\frac{\Gama{-}}{\Gama{1}}$.
Combining these results with Eq.~(6) in the main text and \eqref{sw} gives Eq.~(7) in the main text.

A general solution can be conveniently found by combining \eqref{heom2} with the quantum regression theorem to define the two-time correlation functions, and the Fourier transform can be performed following the approach used in \cite{lu2021characterizing,koshino}.  The result is cumbersome, but taking the strong-driving limit  we obtain the result in Eq.~(8) in the main text.

\section{Three-level master equation}

To evaluate the influence of higher levels  in the transmon on our results, we consider a three-level master-equation model. We describe the energy difference between levels $1$ and $2$ with the parameter $\omega_{12}$, such that the anharmonicity is given by $\delta/2\pi =( \omega_{12}-\omega_{01})/2\pi = -250$ MHz.

To describe the Autler-Townes splitting in the main text we apply two drives at two different frequencies, and in the rotating frame of the two-drive terms we find the system Hamiltonian to be
\beq
\frac{H_S}{\hbar} &=& (\omega_{01}-\omega_{\rm{p}}^{(1)}) |1\rangle\langle 1|+ (\omega_{12}+\omega_{01}-\omega_{\rm{p}}^{(1)}-\omega_{\rm{p}}^{(2)}) |2\rangle\langle 2|+ \frac{\Omega_1}{2}\sigma_x^{(01)} + \frac{\Omega_2}{2}\sqrt{2}\sigma_x^{(12)},
\eeq
where $\omega_{\rm{p}}^{(i)}$ is the drive frequency of input drive $i$, and  we define $\sigma_x^{(ij)} = |i\rangle\langle j|+|j\rangle\langle i|$, $\sigma_-^{(ij)} = |i\rangle\langle j|$, and $\sigma_+^{(ij)} = |j\rangle\langle i|$.

Since we are in a regime where the anharmonicity of the qubit is larger than the decoherence rates \cite{scigliuzzo2020primary}, our master equation is now
\beq
\frac{\partial}{\partial t} \rho_S(t) &=& -\frac{i}{\hbar}[H_S,\rho(t)] \nonumber\\
&+& \frac{\Gama{r}}{2} \left(n_{\rm{r}}^{(01)}+1\right)\mathcal{D}[\sigma_-^{(01)}]\rho(t)
+\frac{\Gama{r}}{2} n_{\rm{r}}^{(01)}\mathcal{D}[\sigma_+^{(01)}]\rho(t)\nonumber\\
&+& \frac{\Gama{n}}{2} \left(n_{\rm{n}}^{(01)}+1\right)\mathcal{D}[\sigma_-^{(01)}]\rho(t) \
+\frac{\Gama{n}}{2} n_{\rm{n}}^{(01)}\mathcal{D}[\sigma_+^{(01)}]\rho(t)\nonumber\\
&+& \Gama{r} \left(n_{\rm{r}}^{(12)}+1\right)\mathcal{D}[\sigma_-^{(12)}]\rho(t)
+\Gama{r}  n_{\rm{r}}^{(12)}\mathcal{D}[\sigma_+^{(12)}]\rho(t)\nonumber\\
&+& \Gama{n} \left(n_{\rm{n}}^{(12)}+1\right)\mathcal{D}[\sigma_-^{(12)}]\rho(t)
+\Gama{n}  n_{\rm{n}}^{(12)}\mathcal{D}[\sigma_+^{(12)}]\rho(t)\nonumber\\
&+&\frac{\Gama{\phi}}{2}\sum_i \mathcal{D}[\ket{i}\bra{i}]\rho(t). \label{3levME}
\eeq

The output power can  be extended to consider two independent coupling operators associated with $\sigma_-^{(01)}$ and $\sigma_-^{(12)}$. Importantly, the $\ket{1}\leftrightarrow\ket{2}$ transition has a  dipole moment which is $\sqrt{2}$ times larger than the $\ket{0}\leftrightarrow\ket{1}$ transition, and which enhances the coupling to the transmission line, increasing both the Rabi drive term and the dissipation rates.

For weak drives on the $\ket{0}\leftrightarrow\ket{1}$ transition alone we can observe that the main influence of the third level is to induce an additional dephasing on the qubit proportional to the thermal excitation rate from the second to the third level.  To evaluate the reflectivity and power spectrum in this case, it is convenient to again work with the Heisenberg equations of motion, where in the limit of $\Omega_2 = 0$ we obtain a closed set of equations for the operators $w_1(t)=\ex{\sigma_-^{(01)}(t)}$, $w_1^*(t)=\ex{\sigma_+^{(01)}(t)}$, $w_2(t) = \ex{\sigma_+^{(01)}(t)\sigma_-^{(01)}(t)}$ and $w_3(t) = \ex{\sigma_-^{(01)}(t)\sigma_+^{(01)}(t)}$, using the normalization condition $ \ex{\sigma_+^{(01)}(t)\sigma_-^{(01)}(t)}+ \ex{\sigma_-^{(01)}(t)\sigma_+^{(01)}(t)}+ \ex{\sigma_+^{(12)}(t)\sigma_-^{(12)}(t)} =1$:

\beq
 \frac{d}{dt} \begin{pmatrix}
  w_1     \\
  w_1^{*} \\
  w_2 \\
  w_3
\end{pmatrix} =
M_2  \begin{pmatrix}
  w_1     \\
  w_1^{*} \\
  w_2\\
  w_3
\end{pmatrix} + B_2 \label{heom3},
\eeq
where
\beq
M_2 =  \begin{pmatrix}
   i\Delta\!-\!\Gamaa{2}{T}&0 &\frac{i\Omega_1}{2} &\frac{-i\Omega_1}{2} \\
   0&-i\Delta\!-\!\Gamaa{2}{T} & \frac{-i\Omega_1}{2} & \frac{i\Omega_1}{2}  \\
   \frac{i\Omega_1}{2} &\frac{-i\Omega_1}{2}  &\! -\! \Gamaa{-}{(01)}\! -\!\Gamaa{1}{(12)} &\Gamaa{+}{(01)}\!-\!\Gamaa{-}{(12)}\\
   \frac{-i\Omega_1}{2} &\frac{i\Omega_1}{2}  &  \Gamaa{-}{(01)}  &\!-\Gamaa{+}{(01)}
\end{pmatrix}\nonumber
\eeq
and
\beq
B_2 = \begin{pmatrix}
  0     \\
  0 \\
 \Gamaa{-}{(12)}\\
 0
\end{pmatrix}
\eeq
and where we use $\Gamaa{2}{T}=\!\Gamaa{2}{(01)}\!+\!\Gamaa{+}{(12)}$, indicating the added dephasing from thermalization of the $\ket{1}\leftrightarrow\ket{2}$ transition, and where $\Gamaa{2}{(01)}=\Gama{\phi}+\Gamaa{1}{(01)}/2$, $\Gamaa{1}{(ij)} = \Gamaa{+}{(ij)}+\Gamaa{-}{(ij)}$,
$\Gamaa{+}{(ij)} = \Gama{r}n_{\rm{r}}^{(ij)}+\Gama{n}n_{\rm{n}}^{(ij)}$, $\Gamaa{-}{(ij)} = \Gama{r}(n_{\rm{r}}^{(ij)}+1)+\Gama{n}(n_{\rm{n}}^{(ij)}+1)$, and $\Delta_{01}= (\omega_{01}-\omega_{\rm{p}}^{(1)})$.  Note that here we have accounted for the larger rates in the $\ket{1}\leftrightarrow\ket{2}$ transitions in \eqref{3levME}, hence parameters like $\Gamaa{+}{(12)}$ are just for convenience, and differ from the actual thermalization rate of the $\ket{1}\leftrightarrow\ket{2}$ transition by a factor of $2$.

{\bf{Reflection coefficient and PSD.}} Evaluating the reflectivity of the $\ket{0}\leftrightarrow\ket{1}$ transition with the above equations, in the limit of weak drive $\Omega_1$, gives
\beq
r = 1- \frac{i\Gama{r}(1-2\Gamaa{+}{(01)}/\Gamaa{1}{(01)})}{\Delta + i(\Gamaa{2}{(01)} + \Gamaa{+}{(12)})}\mathcal{G},
\eeq
where \beq
\mathcal{G}=\frac{\Gamaa{-}{(12)}\Gamaa{1}{(01)}}{(\Gamaa{+}{(01)}\Gamaa{+}{(12)} + \Gamaa{-}{(12)}\Gamaa{1}{(01)})}\approx 1.\eeq

Similarly, for the output power spectrum, we can obtain the strong-drive (large $\Omega_1$) result by generalizing the steps used in the two-level case. For the central peak, around the $\ket{0}\leftrightarrow\ket{1}$ qubit transition frequency, we find


\beq
S(\omega) = \frac{\hbar\omega_{01}\Gama{r}}{2\pi}\frac{\Gamaa{2}{T}}{\left(\omega-\omega_{01}\right)^2+(\Gamaa{2}{T})^2}\mathcal{F},
\eeq
where
\beq
\mathcal{F} = \frac{\Gamaa{-}{(12)}}{2\Gamaa{-}{(12)} + \Gamaa{+}{(12)}} = [\ex{\sigma_+^{(01)}\sigma_-^{(01)}}_{(t\rightarrow \infty)}=\ex{\sigma_-^{(01)}\sigma_+^{(01)}}_{(t\rightarrow \infty)}]_{\Omega_1 \rightarrow \infty}.\nonumber
\eeq
Intuitively, we see here that under a strong drive $\Omega_1$  we still observe the added dephasing ($\Gamaa{2}{T}=\!\Gamaa{2}{(01)}\!+\!\Gamaa{+}{(12)}$), as well as a small change in the steady-state population, on top of the expected result from the two-level model under strong drive, which captures the thermalization with the third level.

For no drive, $\Omega_1 \rightarrow 0$, we find the thermal spectrum as
\beq
S(\omega) = \frac{\hbar\omega_{01}\Gama{r}}{2\pi}\frac{2\Gamaa{2}{T}\Gama{n}\Delta n}{\left(\omega-\omega_{01}\right)^2+(\Gamaa{2}{T})^2}\mathcal{Y},
\eeq
where
\beq
\mathcal{Y} = \frac{\Gamaa{-}{(12)}}{\Gamaa{-}{(12)}\Gamaa{1}{(01)}+\Gamaa{+}{(12)}\Gamaa{+}{(01)}}
\eeq
and $\Delta n = n_{\rm{n}}^{(01)}-n_{\rm{r}}^{(01)}$.

The side-peak contributions to the Autler-Townes data presented in the main text are based on applying a drive on the $\ket{1}\leftrightarrow\ket{2}$ while monitoring the emission from the $\ket{0}\leftrightarrow\ket{1}$ transition. We can also obtain these side-peaks analytically, in a similar fashion as the above calculations.  The Heisenberg equations of motion in this case, for $\Omega_1=0$ and $\Omega_2\neq 0$, form a closed set of equations for the operators $z_1(t)=\ex{\sigma_-^{(12)}(t)}$, $z_1^*(t)=\ex{\sigma_+^{(12)}(t)}$, $z_2(t) \ex{\sigma_+^{(12)}(t)\sigma_-^{(12)}(t)}$, $z_3(t) = \ex{\sigma_-^{(12)}(t)\sigma_+^{(12)}(t)}$,
$z_4 =\ex{\sigma_-^{(01)}(t)}$, $z_4* =\ex{\sigma_+^{(01)}(t)}$, $z_5 = \ex{\sigma_-^{(02)}(t)}$, and $z_5* = \ex{\sigma_+^{(02)}(t)}$.  The full equations of motion and the full result for the output spectrum are cumbersome, but evaluating the side-peaks for the limit of large drive $\Omega_2$ we find
\beq
S(\omega)_{\pm} =\frac{\hbar\omega_{01}\Gama{r}}{2\pi} \frac{2\left(\Gamaa{2}{T}+\Gamaa{2}{(02)}\right)\left(\Gamaa{+}{(01)}+n_{\rm{r}}^{(01)}\left[\Gamaa{1}{(01)} -2\Gamaa{-}{(12)}\right]\right)}{\left(4\left[\omega-\omega_{01}\pm \frac{\Omega_2}{\sqrt{2}}\right]^2 + \left[\Gamaa{2}{T}+\Gamaa{2}{(02)}\right]^2\right)\left(2\left[\Gamaa{-}{(12)}+\Gamaa{+}{(01)}\right]-\Gamaa{-}{(01)}\right)}.
\eeq
Here we introduced  a new parameter which describes the dephasing rate of $z_5$, $\Gamaa{2}{(02)}=\Gama{\phi}+\Gamaa{+}{(01)}/2 + \Gamaa{-}{(12)}$.

If we assume $n_{\rm{n}}^{(01)} = n_{\rm{n}}^{(12)}$ and  $n_{\rm{r}}^{(01)} = n_{\rm{r}}^{(12)}$,  the second term in the numerator simplifies to
\beq
\left(\Gamaa{+}{(01)}+n_{\rm{r}}^{(01)}\left[\Gamaa{1}{(01)} -2\Gamaa{-}{(12)}\right]\right) = 2\Gama{n}\Delta n.
\eeq

The various analytical results obtained in this supplementary material were checked against numerical simulations using QuTiP \cite{qutip, qutip2}.


%

\end{document}